\documentclass[10pt,conference]{IEEEtran}
\IEEEoverridecommandlockouts
\usepackage{graphicx}
\usepackage{url}
\usepackage{amsmath,amssymb,amsfonts}
\usepackage{booktabs}
\usepackage{listings}
\usepackage{multirow}
\usepackage{braket}
\usepackage{ascmac}
\usepackage[T1]{fontenc}

\def\BibTeX{{\rm B\kern-.05em{\sc i\kern-.025em b}\kern-.08em
    T\kern-.1667em\lower.7ex\hbox{E}\kern-.125emX}}

\begin{document}

\title{MAFFT-inspired Quantum Shift-based Sequence Alignment and its Efficient Simulation on Decision Diagrams}

\author{\IEEEauthorblockN{
Yusuke Kimura\IEEEauthorrefmark{1}, Yutaka Takita\IEEEauthorrefmark{1}}
\IEEEauthorblockA{\textit{Fujitsu Limited, Japan}\IEEEauthorrefmark{1}}
\{yusuke-kimura, takita.yutaka\}@fujitsu.com\IEEEauthorrefmark{1}}

\maketitle

\begin{abstract}
Multiple sequence alignment (MSA) is a core operation for comparing genome sequences and is widely used in bio-informatics. MAFFT, a practical MSA tool, repeatedly shifts a pair of sequences and computes a distance. Because the number of sequence pairs grows quadratically with the number of sequences, this procedure can become a bottleneck.

We propose Quantum Shift-based Sequence Alignment (QShift-SA), which implements this ``shift-wise score computation'' as a gate-based quantum circuit and searches over shift amounts and sequence pairs using Grover algorithm. QShift-SA constructs an oracle circuit that compute the Hamming distance (the number of mismatches) between two sequences with data encoding,  controlled shift, comparison, and addition. This oracle can search for candidates with small distances. QShift-SA does not aim to replace the full MSA workflow; instead, it targets the screening steps that often dominate the runtime in classical MAFFT as stated above.

We evaluate circuit resources (number of qubits, gate count, and depth) and benchmark simulation time across multiple quantum circuit simulators. We find that a decision diagram (DD)-based quantum circuit simulator runs more than 1,000$\times$ faster than state-vector and MPS simulators and can handle larger circuits.
\end{abstract}


\section{Introduction}
Next-generation sequencing (NGS) has driven rapid growth in nucleotide sequence data in the life sciences~\cite{bigdata}. These data support a wide range of studies---including precision medicine, drug discovery, and evolutionary biology---but processing large datasets has become increasingly costly.

Multiple sequence alignment (MSA) is a fundamental task in sequence analysis~\cite{mount2004bioinformatics}. It aligns three or more sequences by inserting gaps and reveals similarity at the nucleotide (or amino-acid) level. Computing an optimal MSA exactly is NP-hard~\cite{doi:10.1089/cmb.1994.1.337}, and practical tools therefore rely on heuristics~\cite{progressive,SAGA}. A representative family is progressive alignment~\cite{progressive}, which (i) estimates distances of multiple sequence pairs, (ii) builds a distance-based guide tree (clustering similar sequences first), and (iii) aligns sequences progressively along the tree.

MAFFT~\cite{mafft,mafftv7} is a practical MSA tool based on progressive alignment. It uses the FFT (Fast Fourier Transform) to accelerate the evaluation of match counts (or distances) while shifting two sequences relative to each other, achieving a good balance between speed and accuracy.
The same shift-wise evaluation appears repeatedly, for example, (a) when estimating distances to populate the distance matrix and (b) when detecting similar regions (anchors) for downstream alignment. Even if each pairwise computation is fast, the number of pairs grows as $O(M^2)$ for $M$ sequences, and this step can become a bottleneck~\cite{6883183}.

Quantum computing has been explored as a possible approach to this challenge. Quantum algorithms can accelerate specific problems by exploiting superposition and entanglement~\cite{Shor,Grover}. Additionally, some studies have formulated the entire MSA or guide-tree construction as a QUBO (quadratic unconstrained binary optimization) problem and tackled it using quantum annealing or NISQ optimization (e.g., QAOA)~\cite{qaoa_msa,maq_annealing,qa_guidetree}. While these works provide useful formulations, whether they can yield practical quantum speedups still requires careful assessment. Quantum circuits have also been proposed to compute pairwise sequence similarity with Jaccard distance~\cite{qce23_jaccard}. In contrast, to the best of our knowledge, there is no prior study that designs a concrete gate-based circuit for MSA, and further discusses circuit resources and quantum speedup.

We propose Quantum Shift-based Sequence Alignment (QShift-SA) to target the ``shift-amount search over many pairs,'' which tends to dominate the runtime in MAFFT. QShift-SA (i) stores shift amounts (and data addresses) in binary registers, (ii) uses a circuit that computes the Hamming distance for each shift amount as an oracle, and (iii) searches for promising candidates using Grover search and its extensions such as BBHT~\cite{BBHT} or GAS~\cite{GAS}. QShift-SA does not directly solve gap-aware MSA; rather, it assists classical MAFFT by screening candidates. These are representative use-cases:
\begin{itemize}
\item detect pairs with a distance below a threshold to narrow down candidates for the distance matrix,
\item extract only near neighbors of each sequence to assist guide-tree construction,
\item in iterative workflows, run exact classical DP only for candidates that are likely to have small distances,
\end{itemize}

The main contributions of this work are as follows.
\begin{itemize}
\item Proposing QShift-SA: To implement shift-wise score computation as a quantum circuit, we design a circuit composed of cyclic shifts, comparisons, and Hamming-distance counting, and combine it with Grover search.
\item Resource analysis and bottlenecks: Treating the sequence length $N$ and the number of candidate sequences $M$ as variables, we count the required number of qubits, gate count, and depth, and clarify that state preparation and the shift circuit can become dominant.
\item Simulation characteristics: We simulate the proposed circuit using three approaches ---state-vector (SV), matrix product state (MPS), and decision diagram (DD)--- and show that the DD-based approach achieves more than 1000$\times$ speedup and can handle more qubits.
\end{itemize}

The rest of this paper is organized as follows. Section~II reviews sequence alignment, the relevant component of MAFFT, Grover search, and quantum circuit simulators. Section~III explains our motivation, and Section~IV presents the QShift-SA circuit. Section~V analyzes circuit resources, and Section~VI reports simulation results. Finally, Section~VII concludes the paper.
\section{Preliminaries}

\subsection{Sequence alignment}
Sequence alignment is a fundamental technique for evaluating similarity among biological sequences such as DNA, RNA, and proteins~\cite{mount2004bioinformatics}. A biological sequence can be represented as a string over a finite alphabet $\Sigma$. For DNA sequences, for example, $\Sigma=\{\mathrm{A},\mathrm{C},\mathrm{G},\mathrm{T}\}$.

Given two sequences $X=x_1x_2\dots x_n$ and $Y=y_1y_2\dots y_m$, an alignment inserts a gap symbol (blank) to extend them to the same length and matches characters position by position. Each matched position is classified as a match, mismatch, or gap, and a score (or cost) is assigned so as to maximize the total score (or minimize the total cost).

\subsubsection{Pairwise alignment}
The problem of finding an alignment between two sequences is called pairwise alignment~\cite{NEEDLEMAN1970_pairwise}. As a simple example, consider
\[
X=\mathrm{ACGCT},\quad
Y=\mathrm{TACGC}
\]

\paragraph{Hamming distance}
The Hamming distance $d_H(X,Y)$ is defined, for two strings of the same length, as the number of positions at which the corresponding symbols are different. Comparing the two sequences above position by position gives
\[
\begin{array}{ccccc}
\mathrm{A} & \mathrm{C} & \mathrm{G} & \mathrm{C} & \mathrm{T} \\
\mathrm{T} & \mathrm{A} & \mathrm{C} & \mathrm{G} & \mathrm{C}
\end{array}
\]
and since the symbols differ at all positions, $d_H(X,Y)=5$. While the Hamming distance is easy to compute, it does not allow insertions or deletions (gaps), so it is limited as a similarity measure when shifts or indels must be taken into account.

\paragraph{Edit distance}
The edit distance (Levenshtein distance) $d_E(X,Y)$ is defined as the minimum number of edit operations required to transform one string into the other~\cite{1966Levenshtein}. Allowed operations are insertion, deletion, and substitution. In the example above, we can make them match by inserting $\mathrm{T}$ at the beginning of $X$ and deleting the trailing $\mathrm{T}$:
\[
\begin{array}{cccccc}
- & \mathrm{A} & \mathrm{C} & \mathrm{G} & \mathrm{C} & \mathrm{T} \\
\mathrm{T} & \mathrm{A} & \mathrm{C} & \mathrm{G} & \mathrm{C} & -
\end{array}
\]
Because two operations (one insertion plus one deletion) are required, $d_E(X,Y)=2$. In general, the edit distance between strings of length $n$ and $m$ can be computed by dynamic programming in $O(nm)$ time~\cite{NEEDLEMAN1970_pairwise}.

\subsubsection{Multiple sequence alignment}
The problem of aligning three or more sequences simultaneously is called multiple sequence alignment (MSA)~\cite{mount2004bioinformatics}. One representative objective is the Sum-of-Pairs (SP) score, which sums the scores over all pairs of sequences in the resulting multiple alignment:
\[
SP=\sum_{1\le i<j\le k}\sum_{l=1}^{L} s(a_{i,l},a_{j,l}),
\]
where $k$ is the number of sequences, $L$ is the alignment length, $a_{i,l}$ is the character (including gaps) at position $l$ of sequence $i$, and $s(\cdot,\cdot)$ is a scoring function for two characters. An exact solution that maximizes the SP score can be obtained by high-dimensional dynamic programming, but the computational cost grows exponentially, and MSA is known to be NP-hard~\cite{doi:10.1089/cmb.1994.1.337}.

In practice, heuristics are used. A representative example is the progressive method~\cite{progressive}, which performs MSA in three steps.
\begin{enumerate}
\item Estimate inter-sequence distances to form a distance matrix
\item Build a guide tree (a tree structure that clusters similar sequences first) from the distance matrix
\item Perform MSA progressively along the guide tree
\end{enumerate}
In the distance-estimation stage, many implementations compute a simple score rather than an exact gap-aware score.
Filling the distance matrix requires $\Theta(M^2)$ pairwise distances, and distance estimation itself tends to dominate the runtime for large datasets. Therefore, practical implementations use techniques such as screening candidates with a coarse score or finding only the neighborhood of each sequence.



\subsubsection{Shift-wise score computation in MAFFT}
MAFFT (Multiple Alignment using Fast Fourier Transform) is a fast progressive MSA tool that leverages the FFT and efficiently identifies similar regions (anchors) between sequences~\cite{mafft,mafftv7}. A core subroutine shifts two sequences relative to each other and evaluates their similarity across shift amounts.

This operation ignores gaps and considers only the overlap under each shift, so it does not directly yield a gap-aware alignment. Nevertheless, the shift that produces many matches provides a useful cue to similar regions and serves as a coarse proxy for distance estimation; MAFFT uses this information repeatedly as pre-processing when handling large sequence sets. We briefly review how MAFFT computes these shift-wise scores using the FFT in this section.

\paragraph{Numerical encoding (an example of one-hot representation)}
To apply the FFT, sequences represented as strings must be converted into numerical sequences. Although the encoding is not unique~\cite{mafft}, we map the DNA alphabet $\Sigma=\{\mathrm{A},\mathrm{C},\mathrm{G},\mathrm{T}\}$ to 4-dimensional one-hot vectors for clarity as follows:
\begin{align}
\mathrm{A}=(1,0,0,0), \mathrm{C}=(0,1,0,0)\nonumber\\
\mathrm{G}=(0,0,1,0), \mathrm{T}=(0,0,0,1)\nonumber
\end{align}
Let $\phi:\Sigma\to\mathbb{R}^4$ be the mapping. For sequences $X=x_1x_2\dots x_n$ and $Y=y_1y_2\dots y_n$ (same length $n$), define $u_i=\phi(x_i)$ and $v_i=\phi(y_i)$. Then the (ordinary) inner product $u_i\cdot v_j$ equals 1 when $x_i=y_j$ and 0 otherwise. Using this property, we can count matches.

\paragraph{Number of matches for each shift}
We define the number of matches at shift $k$ as follows.
\begin{equation}
C(k)=\sum_{i=1}^{n} u_i\cdot v_{i+k}
\label{eq:Ck_def}
\end{equation}
A larger $C(k)$ means more matches at that shift. For example, for $X=\mathrm{ACGCT}$ and $Y=\mathrm{TACGC}$, $C(k)$ is maximized at $k=1$.

\paragraph{Batch computation using FFT}
Computing \eqref{eq:Ck_def} directly for each $k$ takes $O(n^2)$ time. MAFFT uses the convolution theorem to compute $C(k)$ for all $k$ collectively via FFT and inverse FFT (IFFT)~\cite{mafft}.
In the one-hot representation, $u_i=(u_{i,\alpha})_{\alpha\in\Sigma}$ and $v_i=(v_{i,\alpha})_{\alpha\in\Sigma}$, and
\[
u_i\cdot v_{i+k}=\sum_{\alpha\in\Sigma}u_{i,\alpha}v_{i+k,\alpha}
\]
therefore,
\[
C(k)=\sum_{\alpha\in\Sigma}\underbrace{\sum_{i=1}^{n}u_{i,\alpha}v_{i+k,\alpha}}_{=:C_{\alpha}(k)}
\]
which can be decomposed as follows. Here, $u_{\cdot,\alpha}$ is a 0/1 sequence indicating whether each position is base $\alpha$, and $C_{\alpha}(k)$ is the number of matches for base $\alpha$. $C_{\alpha}(k)$ can be computed using the Fourier transform $\mathcal{F}$ and its inverse $\mathcal{F}^{-1}$ as follows:
\begin{enumerate}
\item Compute $U_{\alpha}=\mathcal{F}(u_{\cdot,\alpha})$ and $V_{\alpha}=\mathcal{F}(v_{\cdot,\alpha})$
\item In the frequency domain, compute the element-wise product $W_{\alpha}=U_{\alpha}\odot\overline{V_{\alpha}}$ ($\odot$ denotes element-wise multiplication and $\overline{\cdot}$ complex conjugation)
\item Compute $c_{\alpha}=\mathcal{F}^{-1}(W_{\alpha})$; then $c_{\alpha}[k]=C_{\alpha}(k)$
\item Finally, sum them as $C(k)=\sum_{\alpha\in\Sigma}c_{\alpha}[k]$
\end{enumerate}
Because FFT and IFFT can both be computed in $O(n\log n)$ time, the entire vector $C(k)$ can be obtained with the same order of complexity.
In this way, MAFFT accelerates distance computation between two sequences.

\begin{figure}[tb]
    \centering
    \includegraphics[width=80mm]{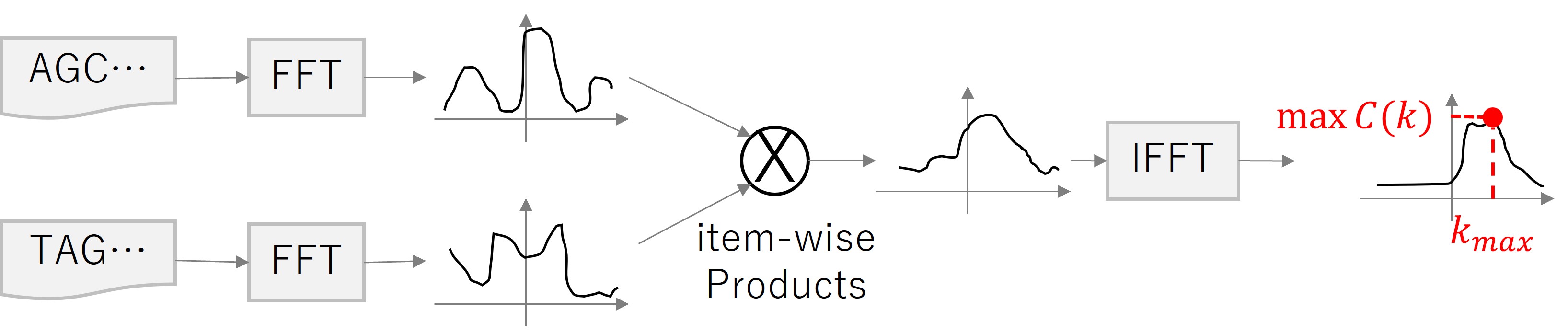}
    \caption{An overview of shift-wise score computation in MAFFT. After numerically encoding two sequences, Fourier transform, element-wise multiplication in the frequency domain, and inverse transform together yield the score vector $C(k)$ for all shift amounts.}
    \label{fig:mafft}
    \begin{minipage}[t]{0.85\linewidth}\footnotesize
    Although we described it here as circular correlation, implementations must choose how to handle boundary conditions, e.g., by approximating non-circular correlation via zero padding.
    \end{minipage}
\end{figure}

\subsection{Grover search}
This subsection briefly reviews Grover search and several extensions referenced in this paper.

\subsubsection{Overview}
Grover search~\cite{Grover} is known to reduce the number of queries for unstructured search problems to roughly the square root of the search-space size. Let the search space size be $N=2^n$, and suppose we are given the following black-box function.
\begin{equation}
f:\{0,1\}^n \rightarrow \{0,1\}
\end{equation}
The goal is to find a solution $x$ such that $f(x)=1$.

\begin{figure}[tb]
    \centering
    \includegraphics[width=80mm]{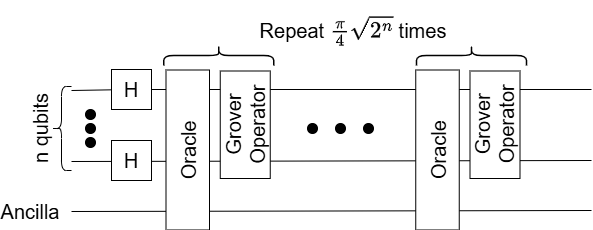}
    \caption{Conceptual diagram of Grover search (repeating the oracle and the diffusion operator)}
    \label{fig:grover}
\end{figure}

In Grover search, we first prepare an $n$-qubit uniform superposition state and then repeatedly apply (i) an oracle that flips the phase of solution states only and (ii) the Grover diffusion operator (inversion about the mean). Repeating these operations amplifies the probability of observing a solution upon measurement. The required number of iterations depends on the number of solutions; when there is a single solution, the algorithm requires about $\frac{\pi}{4}\sqrt{N}$ iterations. Fig.~\ref{fig:grover} illustrates the idea.

\subsubsection{Extensions and discussion}
The standard iteration-count discussion often assumes that the number of solutions is known. In practice, however, the number of solutions may be unknown, or there may be no solutions at all. In such cases, one can use the BBHT algorithm~\cite{BBHT}, which chooses the iteration count to guarantee a bounded success probability, or estimate the number of solutions using Quantum Counting~\cite{Q_Counting} before searching. Grover Adaptive Search (GAS)~\cite{GAS} extends the Grover framework to optimization problems that minimize (or maximize) an objective function.

When applying Grover search to real-data problems, state preparation ---how to load the target data into a quantum state--- can be a bottleneck~\cite{10.1145/3571725,PhysRevX.14.041029}. To address this issue, methods have been proposed to generate Grover oracles more efficiently~\cite{PhysRevA.106.022617,hong2025quantumstatepreparationbased}.
This data-loading mechanism is often referred to as quantum read-only memory (qROM), and there are studies that reduce its circuit depth~\cite{Low_2024,phalak2022optimizationquantumreadonlymemory}.

\subsection{Quantum circuit simulators}
This subsection summarizes three representative types of quantum circuit simulators (SV/MPS/DD) used in our experiments.

\subsubsection{State-vector (SV) simulators}
State-vector simulation is the most standard approach. Because an $N$-qubit quantum state is represented as a $2^N$-dimensional complex vector, the required memory increases exponentially. For example, when complex numbers are stored in double-precision floating point (real and imaginary parts), a 30-qubit state vector requires the following amount of memory.
\begin{equation}
8\times 2 \times 2^{30} = 16\times 2^{30}\;\mathrm{Byte}\approx 16\;\mathrm{GiB}
\end{equation}
State-vector simulators are simple and easy to parallelize, and GPU implementations~\cite{bayraktar2023cuquantum} and multi-node implementations~\cite{imamura2022mpiqulacs} have been proposed. However, the exponential memory requirement in the number of qubits is a fundamental limitation.

\subsubsection{Matrix product state (MPS) simulators}
A matrix product state (MPS) is a representation that expresses a quantum state as a chain of tensors. For qubits, letting $s_i\in\{0,1\}$, it can be written as follows.
\begin{equation}
\ket{\psi}=\sum_{s} \mathrm{Tr}\!\left[A_1^{s_1}A_2^{s_2}\dots A_N^{s_N}\right]\ket{s_1s_2\dots s_N}
\end{equation}
When entanglement is small, the matrix dimensions (bond dimensions) can remain small, enabling memory-efficient and fast simulation. However, as circuits become deeper, the bond dimension tends to grow, and the computational cost can increase rapidly.

\subsubsection{Decision diagram (DD) simulators}
Decision diagrams have long been used to represent logical functions~\cite{10.1109/TC.1986.1676819,Fujita1997}, and they can also be used as compressed representations of quantum states (vectors) and quantum gates (matrices)~\cite{QMDD}. Because a decision diagram (DD) can share identical subgraphs, significant memory reduction and speedup can be expected for quantum states with certain regular structures.

\begin{figure}[tb]
    \centering
    \includegraphics[width=50mm]{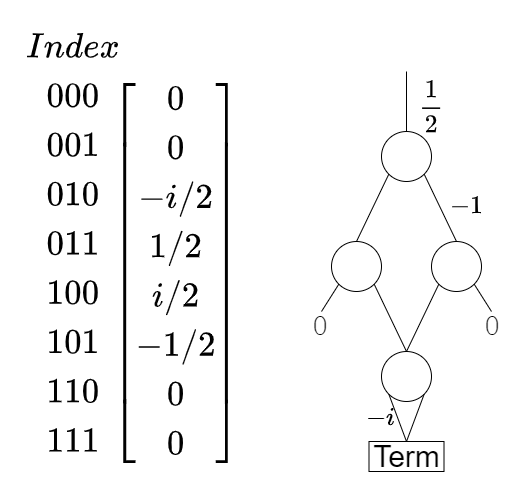}
    \caption{Example of a state-vector representation using a decision diagram}
    \label{fig:dd:vector}
\end{figure}

Fig.~\ref{fig:dd:vector} shows an example of representing a state vector as a decision diagram (DD). By following left/right branches according to the 0/1 bits of the index and multiplying edge weights, the value of each element can be obtained.

Whether a state can be compressed well by a decision diagram (DD) strongly depends on how often identical sub-vectors/sub-matrices repeat. 
DD-based simulators can efficiently handle circuits such as Shor's algorithm or Grover search, whereas they may not be advantageous for circuits with many rotation gates that tend to produce near-random state-vector distributions~\cite{QSW24}.

\section{Motivation}

\subsection{Need for acceleration}
Section~II described how MAFFT shifts two sequences relative to each other and computes the number of matches (or a distance) for each shift amount, using FFT to accelerate the computation for a single pair~\cite{mafft,mafftv7}.

In multiple sequence alignment over $M$ sequences, however, the number of pairs grows as $O(M^2)$, and candidate screening ---deciding which pairs to examine in detail--- can itself become expensive. This occurs, for example, in distance estimation for guide-tree construction and in preprocessing steps that look for similar sequences or anchors. At a high level, the task is a search over candidates (sequence pairs and shift amounts) that satisfy a distance condition.

In $k$-nearest neighbor search (finding only a few close partners per sequence) or filtering (enumerating only pairs whose distance is below a threshold), one does not need the full distance matrix. However, classical pipelines often evaluate many pairs to avoid missing candidates, which scales poorly. This ``few-solution'' approach can be a target for Grover-style search.

Grover search is a representative algorithm for unstructured search and may reduce the number of evaluations to roughly $\sqrt{N}$ for $N$ candidates~\cite{Grover,BBHT}. In our setting, a candidate is specified by (i) the pair of sequences and (ii) the shift amount, yielding about $O(M^2N)$ candidates. Classical computation typically scans this space explicitly, whereas quantum computation can place shift amounts and data addresses in superposition and use distance computation as an oracle to amplify promising candidates.

Grover search is not always beneficial: if preparing the input data as a quantum state is costly, there may be no quantum speedup~\cite{10.1145/3571725,PhysRevX.14.041029}. Therefore, we analyze circuit resources ---including state preparation--- in Sec.~\ref{sec:complexity}.

\subsection{Gap with prior quantum work}
Recent years have seen increasing work on quantum algorithms for sequence analysis~\cite{QCbio_review_mapping,QCcompbio_perspective}. \cite{qce23_jaccard,gasp} conducted pairwise sequence alignment via Hamming/Jaccard distance and combine it with Grover search, and ~\cite{quantum_alignment_partial} proposed circuits that handle a one-character shift against a reference sequence. Another line of work formulates the entire MSA or guide-tree construction as a QUBO and optimizes it using quantum annealing or QAOA~\cite{qaoa_msa,maq_annealing,qa_guidetree}. Our focus differs: we start from MAFFT's ``shift-wise score computation,'' design a concrete gate-based circuit for MSA (multiple sequence alignment), and discuss both circuit resources and quantum speedup.

Finally, simply replacing the classical FFT in MAFFT with the quantum Fourier transform (QFT) is not promising. The QFT is a unitary transform on amplitudes, and unlike the classical FFT, it does not necessarily enable efficiently obtaining all correlation values~\cite{Nielsen_Chuang_2010}. Moreover, MAFFT multiplies FFT outputs element-wise, for which a quantum algorithm is nontrivial. Hence, rather than replacing FFT with QFT, we need a circuit design tailored to this task.

\subsection{Purpose of this study}
Accordingly, we aim to realize the ``shift-wise score computation'' and ``candidate screening'' steps that repeatedly appear in the MAFFT pipeline as a combination of gate-based quantum circuits and Grover search. We construct QShift-SA by (i) designing a circuit that supports arbitrary controlled shifts, (ii) computing distances, and (iii) in the multiple sequence alignment setting, searching over data addresses as well.

\section{Quantum Shift-based Sequence Alignment (QShift-SA)}
\label{sec:qmafft}
In this section, we present QShift-SA, a MAFFT-inspired quantum algorithm that computes Hamming distance under cyclic shifts and searches for promising candidates with Grover search. Each candidate is specified by (i) the pair of sequences and (ii) the shift amount. We store these indices in binary quantum registers to enable Grover search and its extensions. Although this method is also applicable to RNA, amino-acid and etc., this paper uses DNA for easy understanding. 

\subsection{Overview}
For two DNA sequences $X$ and $Y$ of length $N$, let $\mathrm{shift}_k(Y)$ denote the sequence obtained by cyclically shifting $Y$ by $k$. We consider the shift-dependent distance
\begin{equation}
d_H\!\left(X,\mathrm{shift}_k(Y)\right)
\end{equation}
Our goals are (a) thresholded detection: find shift amounts $k$ satisfying $d_H\le \tau$, or (b) minimization: find $k$ that minimizes $d_H$. For multiple sequence alignment, we also search over a candidate set $\{S_i\}_{i=0}^{M-1}$ and seek candidates $(i,j,k)$ with small values of $d_H(S_i,\mathrm{shift}_k(S_j))$.

QShift-SA prepares the shift-amount register and, in the multi-sequence case, the address registers in uniform superposition, applies a distance-computation circuit (controlled shift, comparison, and addition), and then runs Grover search with an oracle that tests whether the distance satisfies the target condition.

\subsection{Register layout}
The basic QShift-SA circuit consists of the following registers. We first describe the pairwise setting, where two sequences of length $N$ are encoded separately.
\begin{itemize}
\item Data registers: store the two sequences. With 2 qubits per base, this requires $4N$ qubits.
\item Shift-amount register: stores $k\in\{0,\dots,N-1\}$ in binary. The required number of qubits is $s=\lceil\log_2 N\rceil$.
\item Comparison and distance registers: store a mismatch bitstring and the accumulated distance (from 0 to $N$). The distance register uses $\lceil\log_2(N+1)\rceil$ qubits.
\item Ancilla registers: used by the comparison circuit and for decomposing MCX gates.
\end{itemize}
For the multiple sequence alignment setting, we additionally introduce two address registers $i$ and $j$ and store $i,j\in\{0,\dots,M-1\}$ in binary (requiring $a=\lceil\log_2 M\rceil$ qubits for each).

\subsection{Data encoding}
The encoding method in this paper is similar to the one used in \cite{qce23_jaccard}.

We encode the DNA alphabet $\{\mathrm{A},\mathrm{T},\mathrm{C},\mathrm{G}\}$ using 2 qubits per base. To load a sequence as a computational-basis state, we apply $X$ gates according to the following mapping:
\[
\mathrm{A}=00,\quad \mathrm{T}=01,\quad \mathrm{C}=10,\quad \mathrm{G}=11
\]

\begin{figure}[t]
  \centering
  \includegraphics[width=0.9\linewidth]{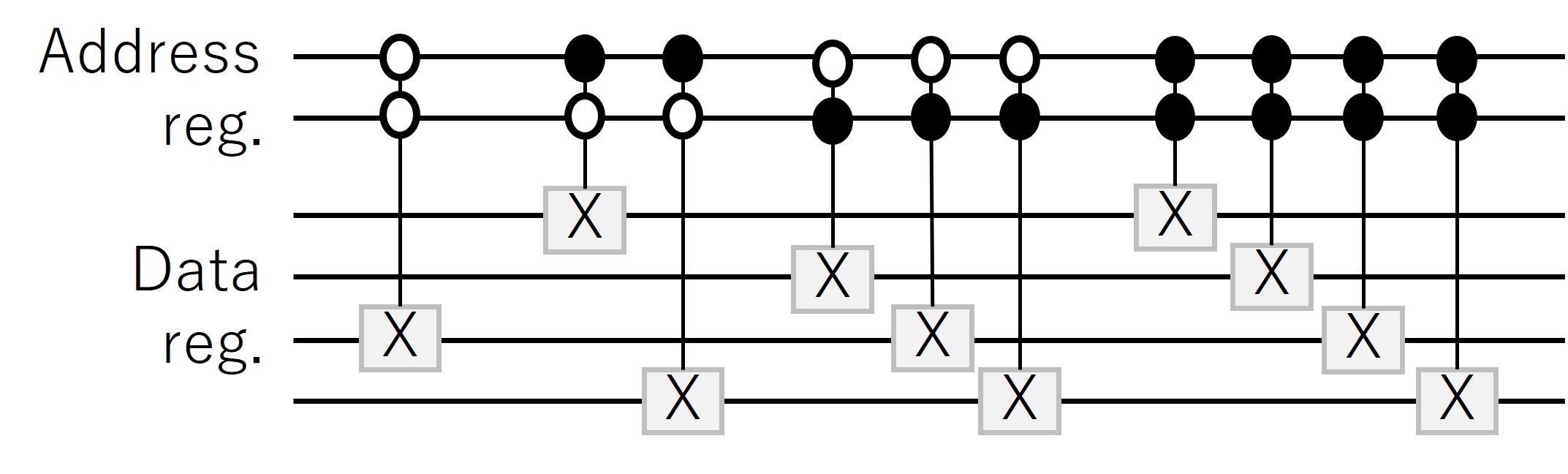}
  \caption{Encoding circuit for four sequences}
  \label{fig:qmafft_multiple}
\end{figure}

The same encoding framework also supports loading multiple candidate sequences in superposition. Using an address register $\lvert a\rangle$ (candidate index) and a data register $\lvert 0\rangle$, we implement a circuit that loads $D[a]$ conditioned on $\lvert a\rangle$ (Fig.~\ref{fig:qmafft_multiple}). This circuit realizes the following transformation.
\[
\sum_a 1/\sqrt{M} \lvert a\rangle \lvert 0\rangle \;\longmapsto\; \sum_a 1/\sqrt{M} \lvert a\rangle \lvert D[a]\rangle
\]
Here, $D[a]$ denotes the $a$-th genome sequence in the dataset.

When the data width $N$ and the number of candidates $M$ are large, the gate count and depth of this encoding circuit can dominate the overall cost, potentially exceeding that of the distance-computation circuit. We revisit this point in Sec.~\ref{sec:complexity}.

\subsection{Controlled cyclic shift circuit}
Classical MAFFT uses FFT/IFFT to identify shift amounts that yield many matches. QShift-SA follows a different route: instead of replacing FFT with QFT, we implement the shift operation directly as a quantum circuit and compute distances on top of it.

\begin{figure}[t]
  \centering
  \includegraphics[width=0.95\linewidth]{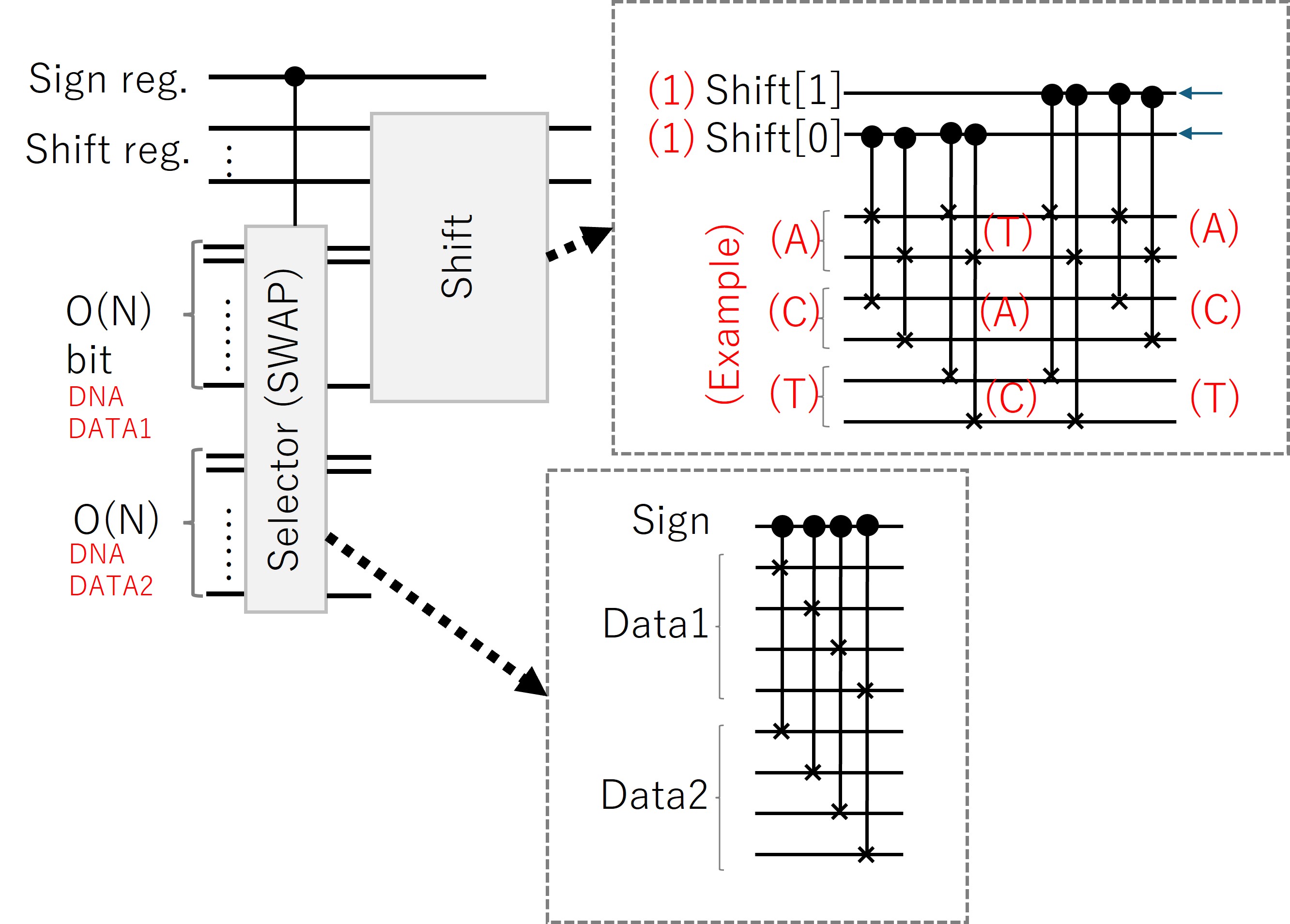}
  \caption{Conceptual diagram of a cyclic shift circuit controlled by the shift-amount register. The shift amount $k$ is represented in binary, and an arbitrary cyclic shift is synthesized by conditionally applying a SWAP network corresponding to each bit.}
  \label{fig:qmafft_shift}
\end{figure}

For simplicity, we use cyclic shifts rather than logical shifts. Because a cyclic shift is a permutation of positions, it can be implemented with a SWAP network. We represent the shift amount $k$ in binary and synthesize the shift in a barrel-shifter style: for each bit $t=0,1,\dots,s-1$, we conditionally apply a SWAP network that shifts by $2^t$ positions when the $t$-th bit of $k$ is 1.
We also include a sign qubit that selects which of the two sequences is shifted.
With this construction, an arbitrary cyclic shift by $k$ can be implemented under the control of the shift-amount register. Fig.~\ref{fig:qmafft_shift} illustrates the idea.

There are several options for constructing the SWAP network. In our implementation, each stage $t$ uses $O(N)$ SWAPs, so the total number of SWAP gates is $O(N\log N)$. Although depth can be reduced with more careful scheduling, we do not optimize it in our implementation.

\subsection{Comparison and distance counting}
\begin{figure}[t]
  \centering
  \includegraphics[width=0.9\linewidth]{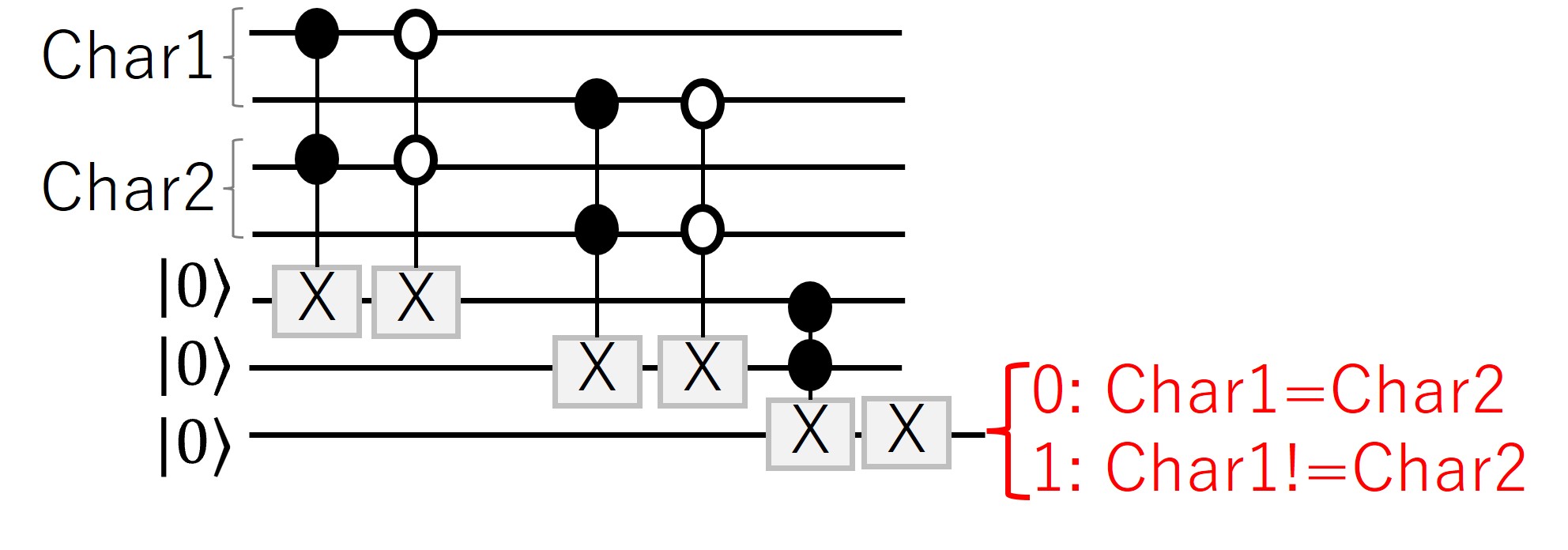}
  \caption{Conceptual diagram of base comparison. The shifted sequence and the reference sequence are compared position by position, producing a mismatch bit for each character.}
  \label{fig:qmafft_compare}
\end{figure}

Each alphabet is encoded in 2 qubits, so we compare the two-qubit symbols position by position and output a mismatch bit that equals 1 when the symbols differ. Concretely, we take XORs of the corresponding bits and then take OR to detect any difference. Fig.~\ref{fig:qmafft_compare} shows a concrete implementation.

Next, we count the number of 1s in the mismatch bitstring and store the result in the distance register. Among several adder designs, we use a QFT-based adder to simplify circuit generation. QFT adders are regular and easy to implement but can be deep; replacing them with a shallower ripple-carry adder or a parallel adder is possible and left for future work.

\begin{figure}[t]
  \centering
  \includegraphics[width=0.95\linewidth]{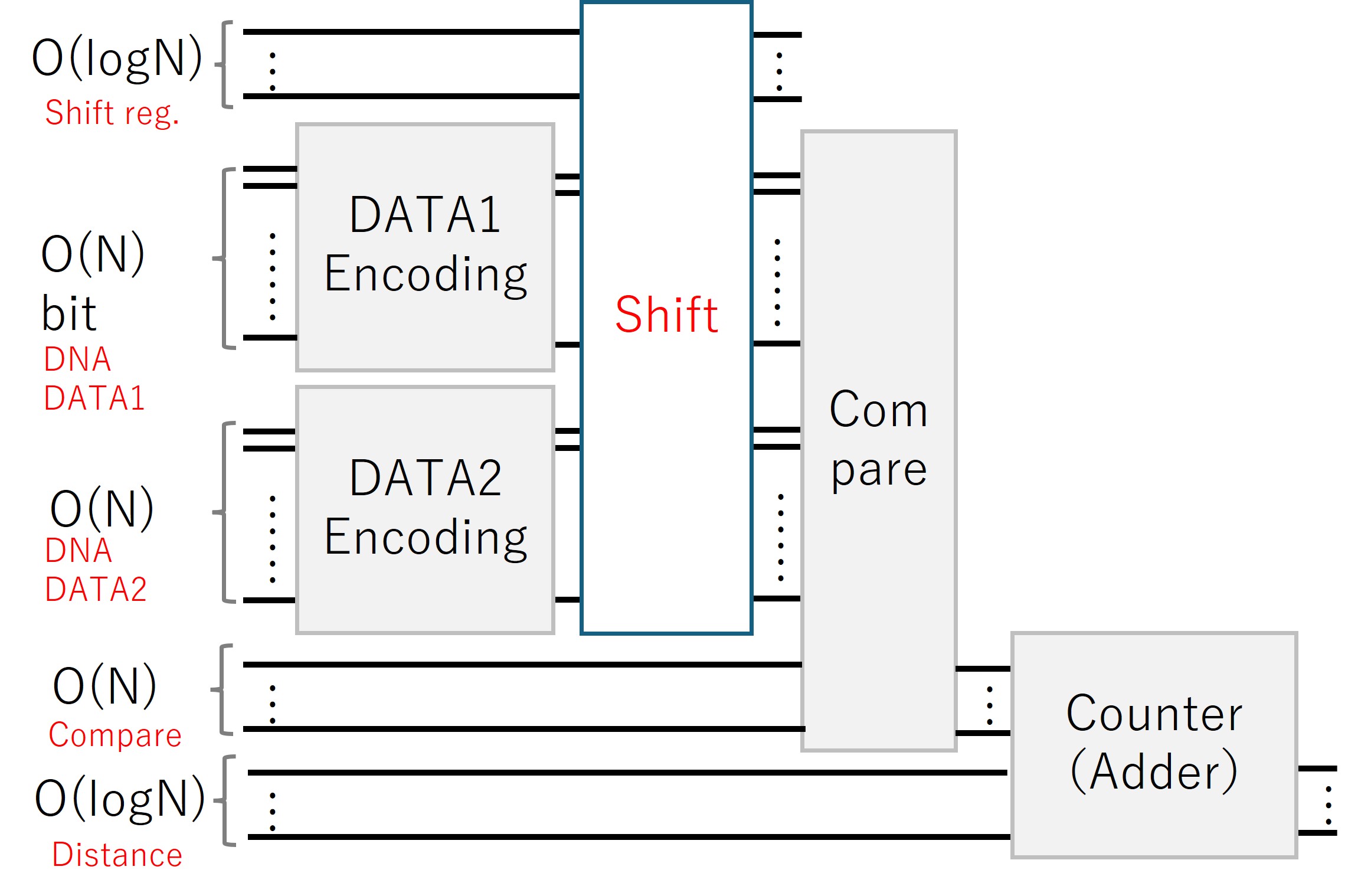}
  \caption{Overview of QShift-SA circuit}
  \label{fig:qmafft_overview}
\end{figure}

Fig.~\ref{fig:qmafft_overview} summarizes the resulting circuit.

\subsection{Combining the oracle with Grover search}
Given the distance-computation circuit, we can build an oracle that flips the phase only for candidate states that satisfy a distance condition. For example, for a threshold $\tau$, we mark states that satisfy
\begin{equation}
1 \le d_H \le \tau
\label{eq:oracle_condition}
\end{equation}
The lower bound ``1'' excludes trivial solutions: if identical sequences are allowed ($i=j$), then $d_H=0$ always holds at $k=0$.



\begin{figure}[t]
  \centering
  \includegraphics[width=0.95\linewidth]{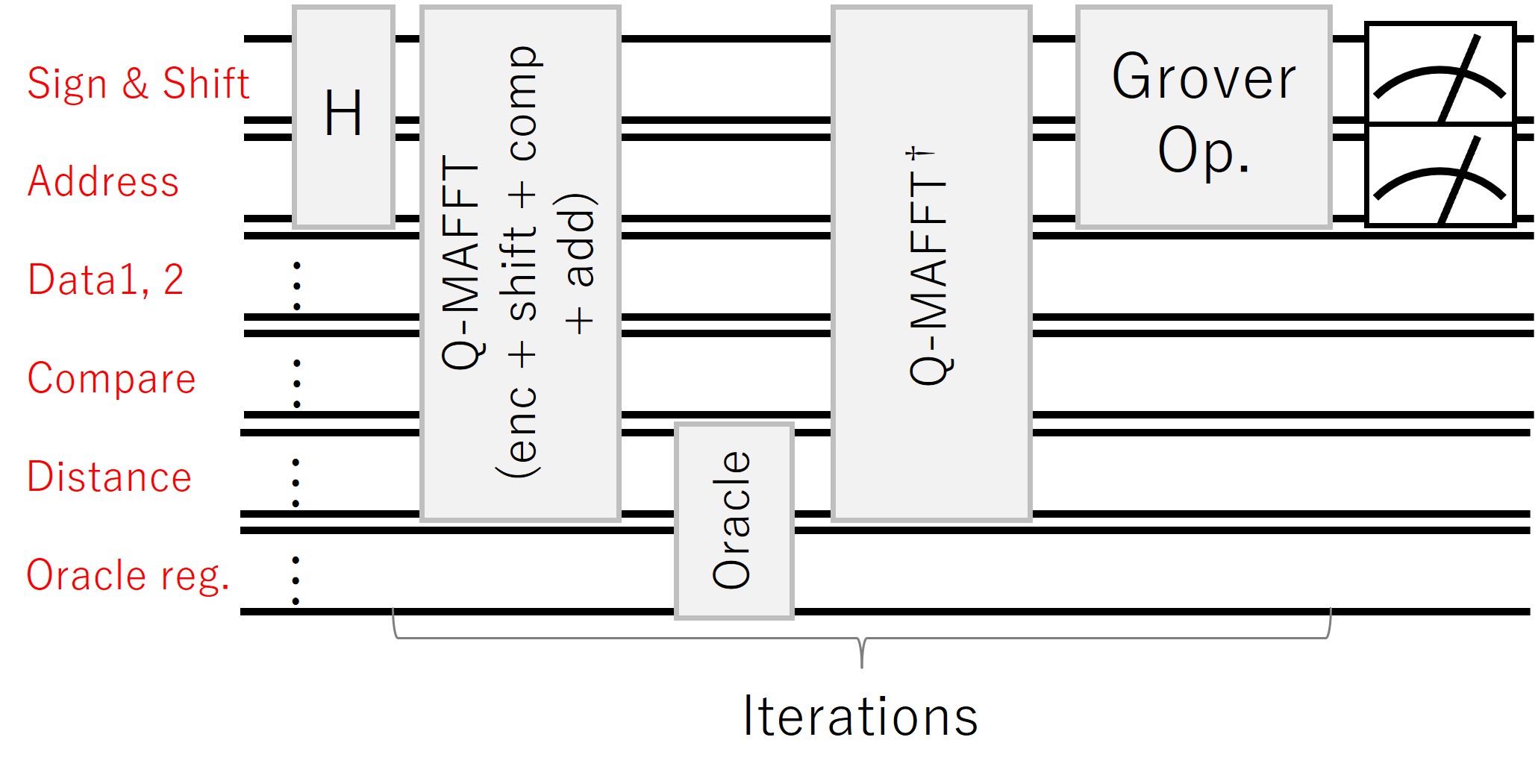}
  \caption{Conceptual diagram of an oracle and Grover search based on a distance condition. We compute $d_H$ using the distance-computation circuit, flip the phase only for candidates that satisfy the condition, and amplify them via the Grover diffusion operator.}
  \label{fig:qmafft_grover}
\end{figure}

By repeating the oracle and the diffusion operator, the amplitude of marked candidates is amplified and they are measured with high probability. Fig.~\ref{fig:qmafft_grover} illustrates the idea.
The search space is represented by the shift-amount register and, in the multiple sequence alignment setting, also by the address registers. We initialize these registers to uniform superposition with Hadamard gates, apply the distance-computation circuit and phase flip, uncompute to reset ancillas, and then apply the diffusion operator. Because the diffusion operator assumes all non-search registers are returned to their initial states, we load data in every iteration and erase it via uncomputation each time.

When the number of marked states is unknown, we can use the BBHT method~\cite{BBHT}. For distance minimization (nearest-neighbor search), we can use frameworks such as GAS~\cite{GAS}, which update the threshold step by step during the search.

We emphasize that QShift-SA is not intended to output the full all-pairs distance matrix. Because the output size is $\Theta(M^2)$, even a quantum algorithm would require at least the same order of measurements to extract all entries. Instead, we use QShift-SA as a substitute for intermediate screening tasks that often dominate MAFFT's runtime, such as searching for close pairs or detecting candidates below a threshold.

In summary, QShift-SA is a quantum circuit that (1) stores addresses and shift amounts in binary, (2) implements a controlled cyclic shift and distance computation, and (3) searches candidates using Grover search (and its extensions). In the next section, we evaluate qubits, gate count, and depth and identify potential bottlenecks.

\section{Complexity Analysis}\label{sec:complexity}
We first discuss, at a high level, when quantum search can be effective in this setting. We then report resource counts in the implementation and discuss bottlenecks.

\subsection{High-level complexity discussion}
\label{subsec:speedup-discussion}
Under a simplified classical cost model, shift-wise score computation for one pair costs about $O(N\log N)$ via FFT/IFFT, and applying it to all pairs yields about $O(M^2N\log N)$. 

The size of the candidate space in QShift-SA is determined by the choice of two sequences and the shift amount and is approximately as follows.
\[
\frac{M(M-1)}{2}\times N = \Theta(M^2N)
\]
With Grover search, the number of oracle calls can scale as the square root of the candidate space; ideally, one might obtain candidates with about $O(\sqrt{M^2N})=O(M\sqrt{N})$ iterations. To realize this benefit, (i) the number of candidates satisfying the condition should not be extremely large (too many solutions weakens amplitude amplification) and (ii) the oracle cost (distance computation plus state preparation) should not dominate.


For (ii), in this paper we use the state-preparation circuit as in Fig.~\ref{fig:qmafft_multiple}, whose gate count grows roughly proportional to $O(MN)$. This differs from the ``low-cost memory access'' assumption often implicit in idealized qROM models and matters whether quantum search is advantageous. Therefore, Table~\ref{tab:submodule_breakdown_cases} separately reports state-preparation and distance-computation costs so that we can see which becomes dominant.

\subsection{Evaluation of implemented circuits}
\subsubsection{How we counted resources}
We report the results of implementing the proposed QShift-SA in Qiskit (v2.2.2). The circuits include high-level gates like MCX (multi-controlled-X) and were transpiled to the desired basis-gate set. We always used optimization level 1.

The primary evaluation target is one Grover iteration. In addition, we measured the same resource metrics for each subcircuit (encoding\_multiple, shift, compare, adder, qshift-sa, diffuser) under the same conditions. We used a simple oracle that flips the phase when the distance equals 1. In Grover iterations, ancilla registers must be returned to the initial state before applying the diffusion operator; thus, distance computation and encoding generally require uncomputation, and the iteration has the structure ``compute + oracle + uncompute + diffusion.'' We count the entire structure as one iteration.

The main results are summarized in Table~\ref{tab:overall_1iter_ucx} (the entire Grover iteration), Table~\ref{tab:overall_1iter_ucx_mcx} (without decomposing MCX gates), and Table~\ref{tab:submodule_breakdown_cases} (subcircuit breakdown).
We used the basis gate set $\{\mathrm{u},\mathrm{cx}\}$, and we also prepared results for $\{\mathrm{u},\mathrm{cx},\mathrm{mcx}\}$, where MCX gates are kept as logical gates without decomposition. Although the u\_cx\_mcx counts do not represent physical execution cost directly, they are useful for isolating how much MCX decomposition increases depth and gate count.

\subsubsection{About one Grover iteration}
Table~\ref{tab:overall_1iter_ucx} shows the resources for one Grover iteration in the $\{\mathrm{u},\mathrm{cx}\}$ basis. As $N$ increases, the number of qubits, depth, and total gate count increase. Even with $N$ fixed, increasing $M$ increases depth and total gates similarly, indicating the strong impact of state preparation. MCX synthesis with ancilla qubits (with\_ancilla) reduces depth and total gates by about 7\% for small-to-medium circuits, but the difference becomes small for larger circuits.

\begin{table}[t]
\centering
\caption{Resources of one Grover iteration (basis $\{\mathrm{u},\mathrm{cx}\}$; parentheses show with\_ancilla)}
\label{tab:overall_1iter_ucx}
\footnotesize
\begin{tabular}{rrrr}
\toprule
$(N,M)$ & Qubits & Depth & Total gates \\
\midrule
$(4,4)$   & 33 (34)   & 1,188 (1,107)   & 2,567 (2,411) \\
$(8,4)$   & 55 (56)   & 2,361 (2,182)   & 5,450 (5,177) \\
$(8,8)$   & 57 (58)   & 5,037 (4,556)   & 10,957 (10,267) \\
$(16,16)$ & 101 (102) & 33,658 (33,614) & 84,558 (84,480) \\
\bottomrule
\end{tabular}
\end{table}

Table~\ref{tab:overall_1iter_ucx_mcx} provides reference values when MCX gates are not decomposed (u\_cx\_mcx). In u\_cx\_mcx, depth and total gates are smaller because each MCX remains as a single gate. For example, for $(16,16)$, the total gate count increases by about 10$\times$ from 8,385 to 84,558 when MCX decomposition is applied, showing that decomposing MCX gates strongly affects the overall cost.

\begin{table}[t]
\centering
\caption{Reference values when MCX gates are kept without decomposition}
\label{tab:overall_1iter_ucx_mcx}
\footnotesize
\begin{tabular}{rrrrr}
\toprule
$(N,M)$ & Qubits & Depth & Total gates & \#MCX \\
\midrule
$(4,4)$   & 33  & 975   & 2,257 & 2 \\
$(8,4)$   & 55  & 1,969 & 4,920 & 2 \\
$(8,8)$   & 57  & 1,437 & 3,382 & 230 \\
$(16,16)$ & 101 & 3,259 & 8,385 & 990 \\
\bottomrule
\end{tabular}
\end{table}

\subsubsection{Subcircuit breakdown and bottlenecks}
\begin{figure}[t]
  \centering
  \includegraphics[width=0.8\linewidth]{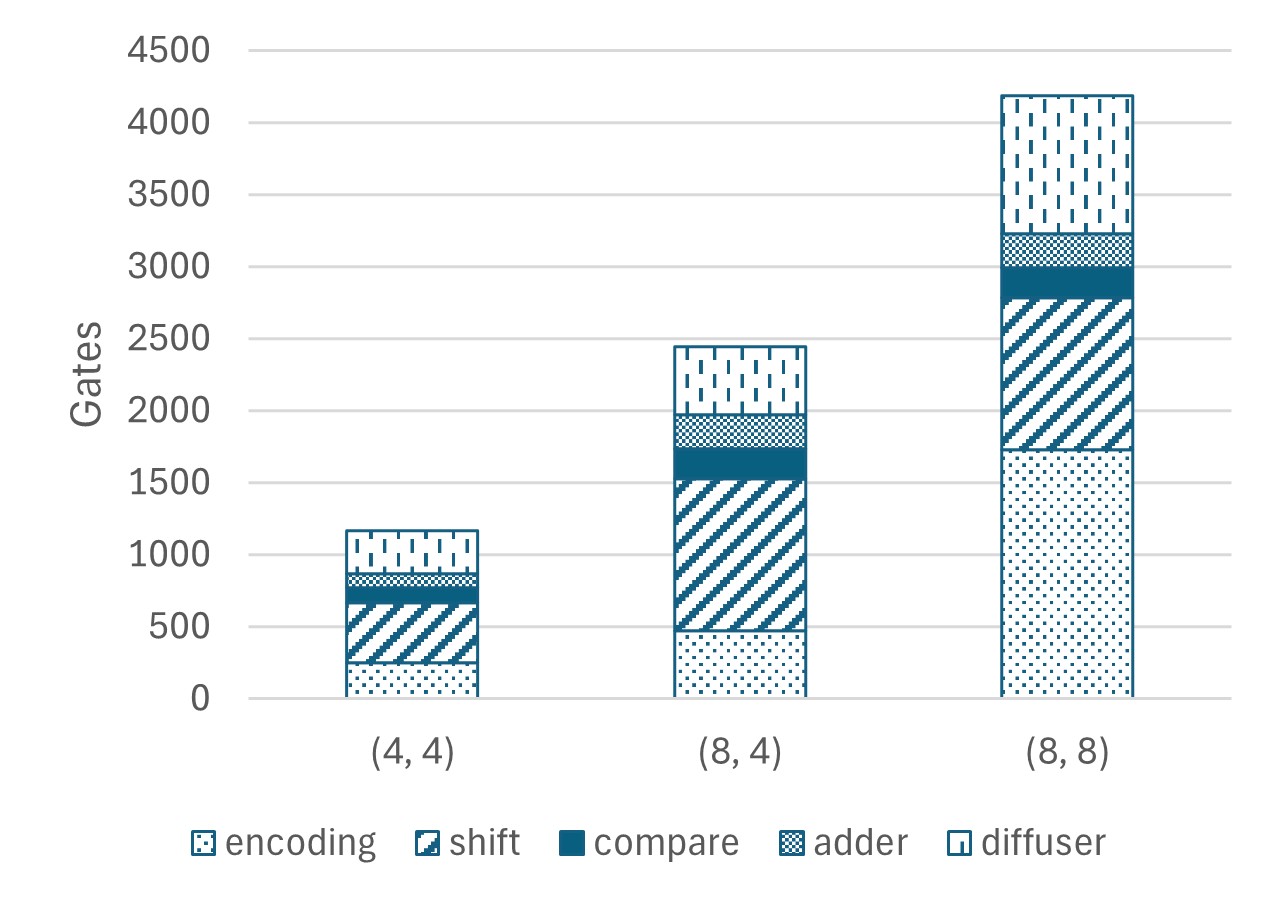}
  \includegraphics[width=0.8\linewidth]{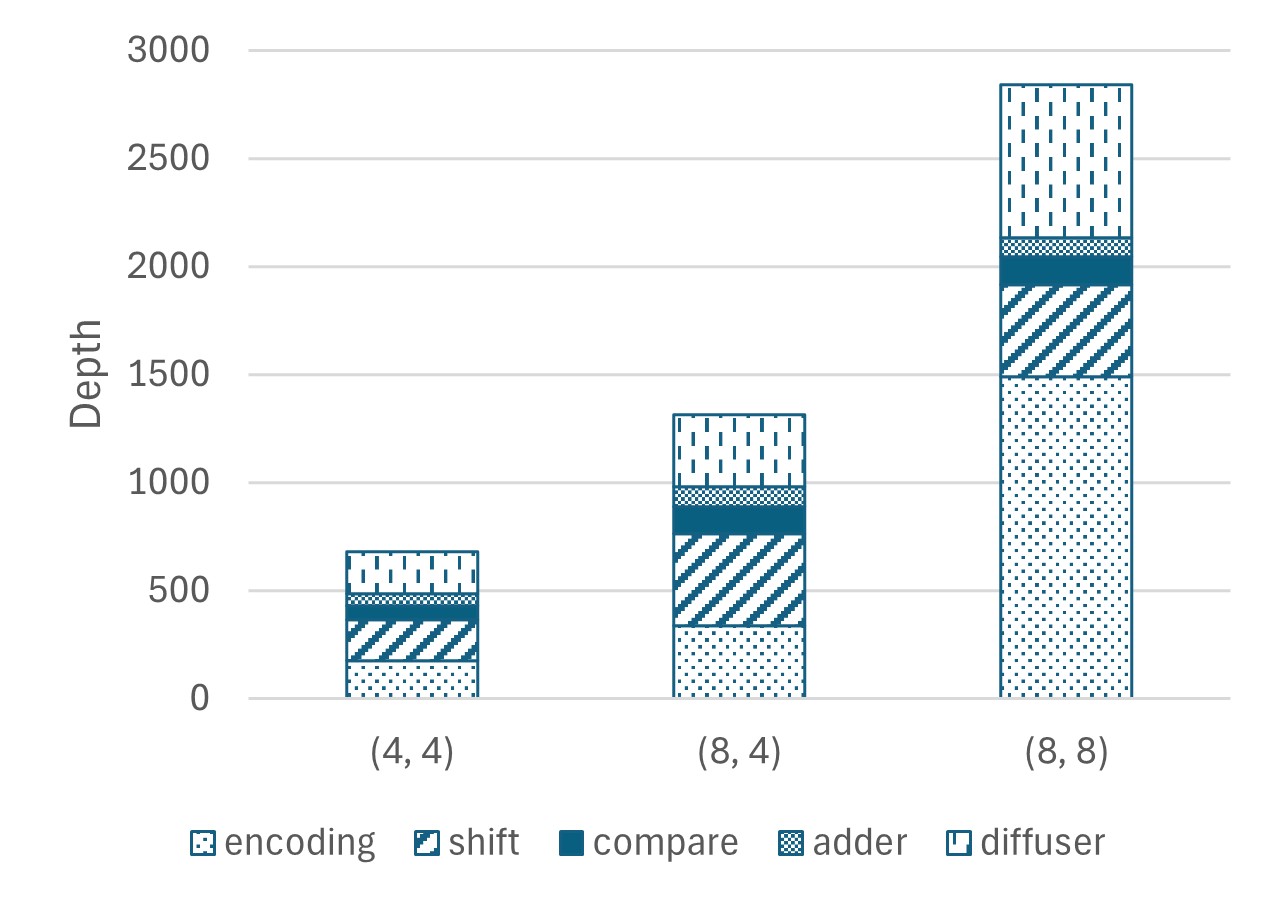}
\caption{Gate count and depth of one Grover-iteration circuit (basis $\{\mathrm{u},\mathrm{cx}\}$)}
  \label{fig:circuit_stats}
\end{figure}

Table~\ref{tab:submodule_breakdown_cases} shows the breakdown of the overall cost. Fig.~\ref{fig:circuit_stats} shows how gate count and depth scale. On the small-scale side (e.g., $(8,4)$), qshift-sa is the main contributor, and within it, shift dominates (1,056 gates out of 1,501 gates in qshift-sa). This indicates that the controlled cyclic shift determines the size of the QShift-SA core.

In contrast, in $(16,16)$, state preparation encoding\_multiple reaches 18,852 gates and depth 14,669, far exceeding qshift-sa (3,523 gates and depth 1,350). That is, when $M$ is large, preparing the candidate set as a quantum state tends to become the bottleneck, rather than distance computation (shift/compare/adder).

Although the QFT adder used for distance counting tends to be deep, shift is the dominant component inside qshift-sa, and replacing the adder alone yields limited improvement. Therefore, the implementation priorities are (i) improving state preparation, (ii) reducing the depth of the shift circuit, and (iii) improving the adder, in this order.

\begin{table*}[t]
\centering
\caption{Total gate count and depth by subcircuit (basis $\{\mathrm{u},\mathrm{cx}\}$, no\_ancilla)}
\label{tab:submodule_breakdown_cases}
\footnotesize
\begin{tabular}{lrrrrrrrr}
\toprule
& \multicolumn{2}{c}{$(4,4)$} & \multicolumn{2}{c}{$(8,4)$} & \multicolumn{2}{c}{$(8,8)$} & \multicolumn{2}{c}{$(16,16)$} \\
Module & Gates & Depth & Gates & Depth & Gates & Depth & Gates & Depth \\
\midrule
encoding\_multiple & 250 & 177 & 474 & 337 & 1,730 & 1,490 & 18,852 & 14,669 \\
qshift-sa            & 619 & 308 & 1,501 & 645 & 1,501 & 645 & 3,523 & 1,350 \\
\quad shift        & 416 & 188 & 1,056 & 425 & 1,056 & 425 & 2,592 & 950 \\
\quad compare      & 104 & 64  & 208  & 128 & 208  & 128 & 416  & 256 \\
\quad adder        & 96  & 56  & 234  & 92  & 234  & 92  & 512  & 144 \\
diffuser          & 300 & 193 & 475 & 333 & 960 & 705 & 1,926 & 1,501 \\
\bottomrule
\end{tabular}
\end{table*}

\section{Experimental Evaluation}

\subsection{Experimental setup}
We ran experiments in a virtualized environment (WSL2, Ubuntu 24.04) on a Windows 11 laptop with an Intel Core i7-1370P CPU and 32~GB memory (16~GB allocated to the virtual environment). The software stack was Python 3.11.10, Qiskit 2.2.2, Qiskit Aer 0.17.2~\cite{aer}, and qdd 0.2.7~\cite{QSW24}.
We used the short DNA sequences from the Rosalind~\cite{rosalind}, which is widely used for educational practice problems.

For quantum circuit simulation, we compared (i) Aer's state-vector backend (SV), (ii) Aer's matrix product state backend (MPS), and (iii) the decision diagram backend (DD) implemented in qdd. For each backend, we generated the same logical circuit in Qiskit, transpiled it to the supported basis, and executed it. We used optimization level 1 for transpilation and fixed the random seed.
The runtime and accuracy of the MPS backend depend on settings such as bond-dimension truncation, and we used Aer's default setting (no bond-dimension limit).

We measured runtime as wall-clock time from starting the backend until the result was returned, excluding circuit generation and transpilation. Each measurement was terminated at one hour; when the limit was exceeded, we recorded it as $>3600$.

\subsection{Two-sequence example: classical FFT score vs. QShift-SA distance}
We verify on a short-sequence case study that
(1) the match score computed by classical FFT/IFFT (the number of matches for each shift), and
(2) the Hamming distance (number of mismatches) computed by the QShift-SA circuit,
are consistent for each shift amount.

\begin{figure}[t]
  \centering
  \includegraphics[width=\linewidth]{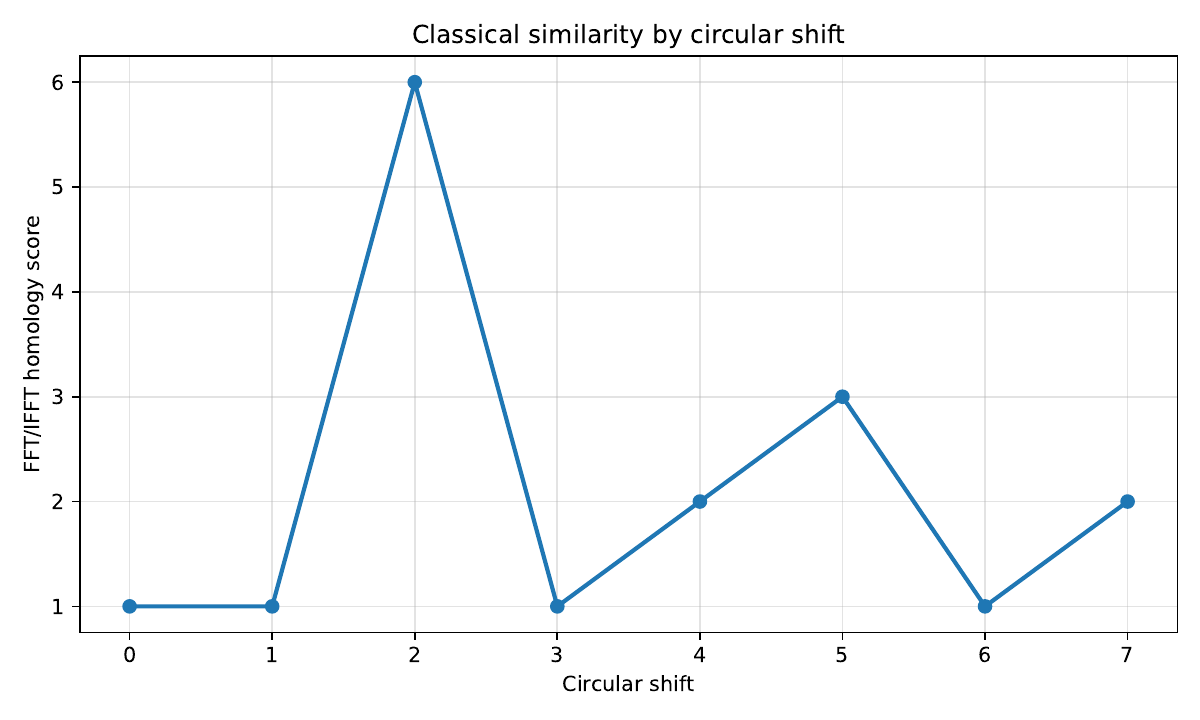}
  \caption{Shift dependence of the match score computed by FFT/IFFT (circular correlation) ($S_0=\mathrm{ATGCAACT}$, $S_1=\mathrm{GCAACTCC}$). Under our definition (shifting $S_1$), the score is maximized at shift=2.}
  \label{fig:fft_similarity_0_1_circular}
\end{figure}

We consider two sequences $S_0=\mathrm{ATGCAACT}$ and $S_1=\mathrm{GCAACTCC}$ (length $N=8$). As a classical MAFFT, we computed the match score using FFT/IFFT; the results are shown in Fig.~\ref{fig:fft_similarity_0_1_circular}. In this setting, the similarity score is maximized at shift=2.

\begin{figure}[t]
  \centering
  \includegraphics[width=\linewidth]{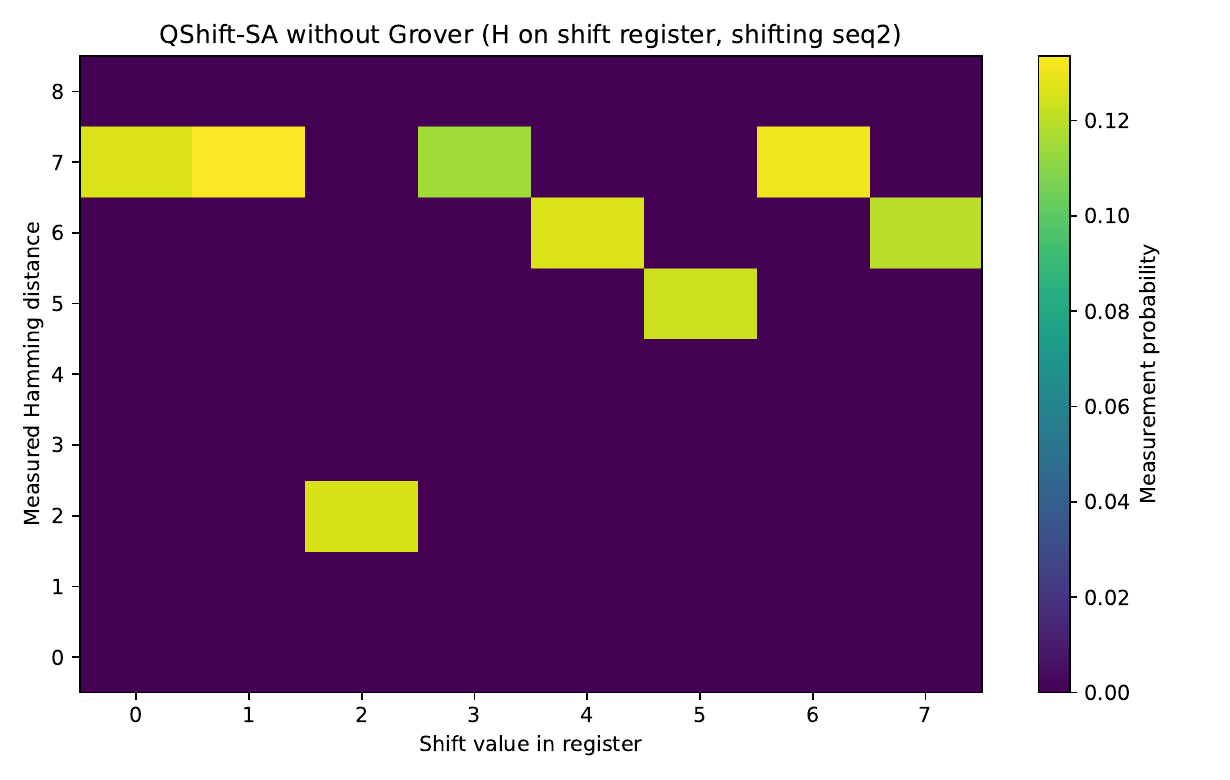}
  \caption{Measured (shift, distance) distribution obtained by QShift-SA without Grover search (uniform superposition over the shift register) ($S_0=\mathrm{ATGCAACT}$, $S_1=\mathrm{GCAACTCC}$). Distance 2 is obtained at shift=2, matching the best shift in classical FFT/IFFT.}
  \label{fig:qmafft_h_shift_0_1_target0}
\end{figure}

Next, without Grover search, we confirm that the distance-computation part of QShift-SA (shift/compare/adder) matches classical computation. We put the shift register into a uniform superposition using Hadamard gates, apply QShift-SA once, and measure (shift, distance); the distribution is shown in Fig.~\ref{fig:qmafft_h_shift_0_1_target0}. For each shift value, only the corresponding distance is observed; in particular, the minimum distance occurs at shift=2, consistent with the FFT/IFFT result above.
Because this experiment does not use the oracle or the diffusion operator, the shift values themselves remain uniformly distributed.

\subsection{Multi-sequence example with Grover}
Finally, as a small-scale example that mimics multiple alignment, we use the following candidate set of sequences (4 sequences of length 8) and run Grover search over a search register consisting of two 2-bit addresses and a 4-bit shift including a sign bit.
\[
[\mathrm{ATGCAACT},\mathrm{GCAACTCC},\mathrm{ACGTTAAA},\mathrm{GTATGCAA}]
\]
In the oracle circuit, we flip the phase only for candidates $(i,j,k)$ whose distance equals 1. Exhaustive classical search yields four solutions, and we set the number of Grover iterations to 6.

\begin{figure}[t]
  \centering
  \includegraphics[width=\linewidth]{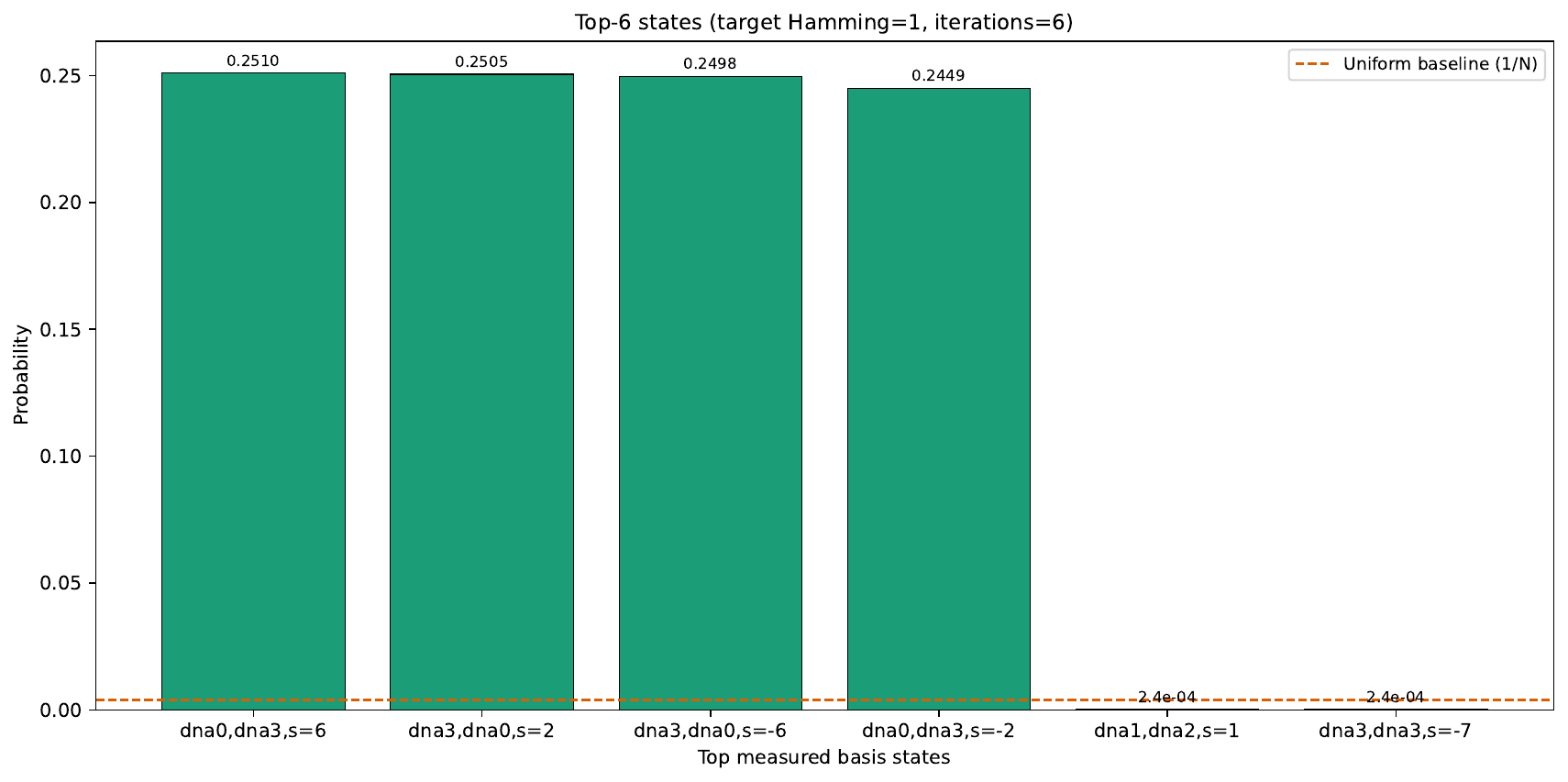}
  \caption{An integrated example of QShift-SA with Grover search (4 candidates, search space size 256, distance=1 as solutions, 6 Grover iterations, 4096 shots).}
  \label{fig:grover_hamming_h1_it6_topk}
\end{figure}

Fig.~\ref{fig:grover_hamming_h1_it6_topk} shows the measurement results (top six states) from 4096 shots. Solution states are observed with high probability, showing a strong bias compared with the uniform probability $1/2^8\approx 3.9\times 10^{-3}$. This indicates that the QShift-SA distance computation and oracle functioned correctly within the Grover search loop. The four solutions are equivalent considering the direction of cyclic shifts and the ordering of sequence pairs.

\subsection{Simulation performance comparison (SV/MPS/DD)}
\begin{figure}[t]
  \centering
  \includegraphics[width=0.8\linewidth]{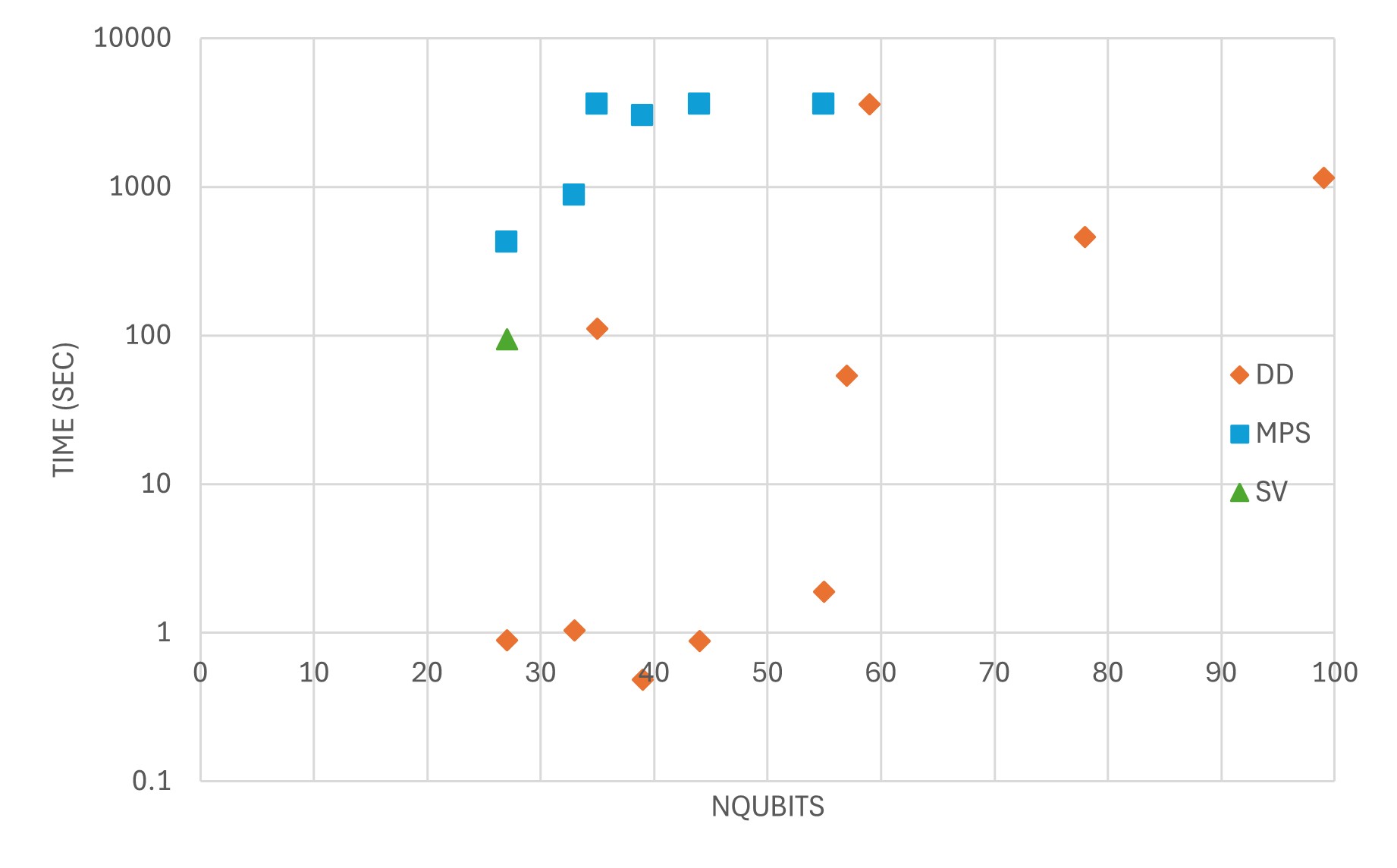}
  \caption{Simulation runtime of the QShift-SA+Grover circuit (log scale). $>3600$ indicates that the measurement was terminated at 3600~s.}
  \label{fig:sim_runtime_vs_qubits}
\end{figure}

We compare the simulation runtime of the QShift-SA (Grover-integrated) circuit across three methods: (i) SV, (ii) MPS, and (iii) DD, and discuss the observed performance differences. Fig.~\ref{fig:sim_runtime_vs_qubits} plots runtime versus the number of qubits on a log scale, and Table~\ref{tab:sim_time_summary} summarizes the key results.

\begin{table}[t]
\centering
\caption{Summary of simulation time (s). Due to memory constraints, SV runs only up to roughly 30 qubits; rows beyond that were not executed and are shown as ``--''.}
\label{tab:sim_time_summary}
\footnotesize
\begin{tabular}{rrrrrr}
\toprule
$N$ & $M$ & \#Qubits & DD & MPS & SV \\
\midrule
3 & 4 & 27 & 0.89 & 422.40 & 93.89 \\
4 & 4 & 33 & 1.04 & 877.04 & -- \\
5 & 4 & 39 & 0.48 & 3,000.47 & -- \\
6 & 4 & 44 & 0.88 & $>3600$ & -- \\
8 & 4 & 55 & 1.88 & $>3600$ & -- \\
4 & 8 & 35 & 110.89 & $>3600$ & -- \\
8 & 8 & 57 & 53.18 & $>3600$ & -- \\
12 & 8 & 78 & 459.49 & $>3600$ & -- \\
16 & 8 & 99 & 1,142.57 & $>3600$ & -- \\
8 & 16 & 59 & $>3600$ & $>3600$ & -- \\
\bottomrule
\end{tabular}
\end{table}

Fig.~\ref{fig:sim_runtime_vs_qubits} and Table~\ref{tab:sim_time_summary} show that DD-based simulation is faster over a wider range and reaches larger circuits than the other methods. For 27 qubits, DD finishes in 0.89~s, whereas SV takes 93.89~s and MPS takes 422.40~s; DD is about 100 times faster than SV and about 500 times faster than MPS.

However, even with DD, runtime can exceed 3600~s depending on the setting. As shown in Table~\ref{tab:submodule_breakdown_cases}, when $M$ is large, the state-preparation circuit becomes huge and dominates simulation time ($N=8,M=16$).

The speed of the DD approach stems from the fact that the QShift-SA circuit contains few controlled rotation gates and yields substantial node sharing in the decision diagram. In contrast, SV simulation requires $2^N$ memory for $N$ qubits, which limited our runs to at most 30 qubits. In the MPS approach, many multi-qubit operations act across distant qubits, producing substantial entanglement and leading to longer simulation time.

In conventional quantum genomics research, DD-type simulators were not utilized, limiting algorithm verification to extremely small problem scales. Thanks to the utility of DD-type quantum circuit simulators demonstrated in this study, it is expected that genome analysis using quantum algorithms will gain broader attention.

\section{Conclusion}

Inspired by a subroutine in classical MAFFT ---evaluating match counts or distances while shifting two sequences--- we proposed QShift-SA, a quantum algorithm that combines this computation with Grover search to find promising candidates. QShift-SA is not a replacement for the full MSA workflow; rather, it targets ``shift-amount search / candidate screening over many pairs,'' which tends to become expensive in classical MAFFT.

In the experiments, we evaluated qubits, gate count, and depth via circuit counting and showed that state preparation can become a bottleneck in the multiple sequence alignment setting. We also compared simulation performance across SV/MPS/DD and confirmed that the DD-based approach is substantially faster.

Our evaluation is based on classical simulation, and we do not claim that circuits of this scale can be executed on real quantum hardware today. However, explicitly identifying which components grow rapidly and which assumptions (e.g., data encoding) are critical to potential speedups is useful for future implementations and method comparisons.

Future work includes extending the method to logical shifts and to more biologically plausible scoring schemes. Our experiments also indicate that state preparation becomes a bottleneck when the number of sequence large. It would be promising to leverage methods that generate Grover oracles more efficiently~\cite{PhysRevA.106.022617,hong2025quantumstatepreparationbased} and methods that reduce the circuit depth of qROM down to $\sqrt{NM}$~\cite{Low_2024,phalak2022optimizationquantumreadonlymemory}.

\bibliographystyle{IEEEtran}
\bibliography{main}
\end{document}